\documentclass[prl,amsmath,amssymb,aps,superscriptaddress,floatfix,reprint,notitlepage]{revtex4-1}
\usepackage{times}
\usepackage{graphicx}
\usepackage{amsmath, braket, amsfonts}
\usepackage{amssymb}
\usepackage{natbib}
\usepackage{sidecap}
\usepackage{bm, color, ulem}
\usepackage{xcolor}
\usepackage{lipsum}
\DeclareUnicodeCharacter{03BD}{{$\nu$}}

\begin{document}

\title{Superconductivity in rhombohedral trilayer graphene}
\author{Haoxin Zhou}
\affiliation{Department of Physics, University of California at Santa Barbara, Santa Barbara CA 93106, USA}
\author{Tian Xie}
\affiliation{Department of Physics, University of California at Santa Barbara, Santa Barbara CA 93106, USA}
\author{Takashi Taniguchi}
\affiliation{International Center for Materials Nanoarchitectonics,
National Institute for Materials Science,  1-1 Namiki, Tsukuba 305-0044, Japan}
\author{Kenji Watanabe}
\affiliation{Research Center for Functional Materials,
National Institute for Materials Science, 1-1 Namiki, Tsukuba 305-0044, Japan}
\author{Andrea F. Young}
\email{andrea@physics.ucsb.edu}
\affiliation{Department of Physics, University of California at Santa Barbara, Santa Barbara CA 93106, USA}
\date{\today}
\begin{abstract}
\end{abstract}
\maketitle

\begin{figure*}[t]
\centering
\includegraphics[width=\textwidth]{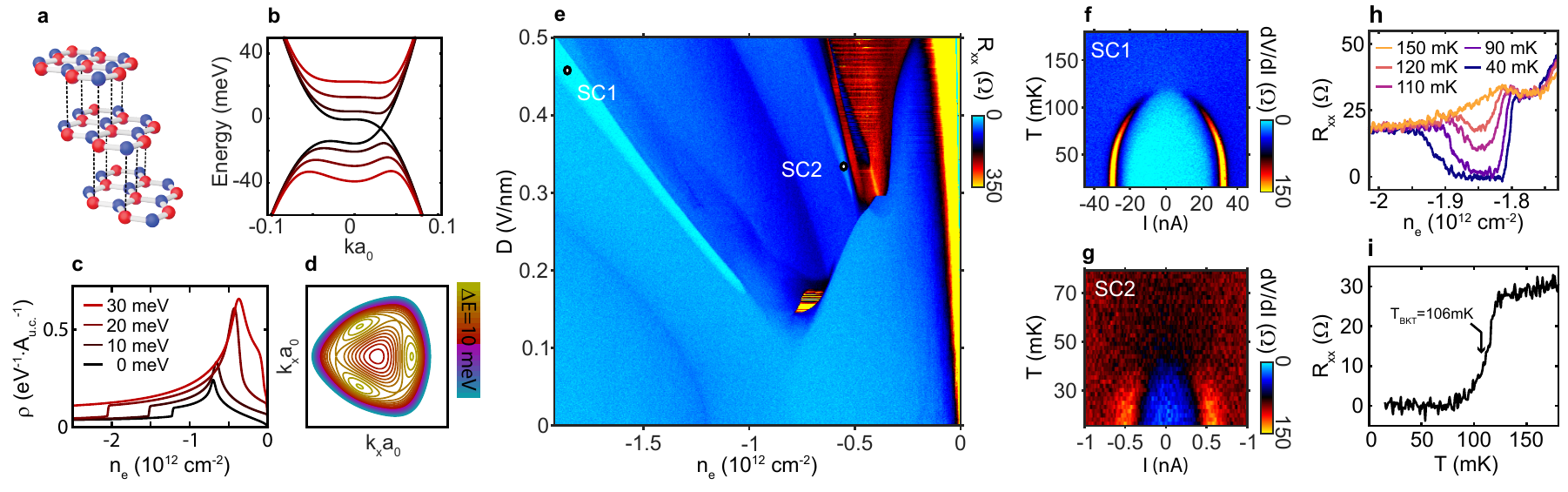}
\caption{\textbf{Superconductivity in rhombohedral trilayer graphene (RTG).}
\textbf{a}, Crystal structure of RTG.
\textbf{b}, Band structure of RTG for interlayer potential $\Delta_1$=0, 10, 20, and 30 meV. 
\textbf{c}, Density of states, $\rho$ calculated in the single particle model.
\textbf{d} Isoenergetic contours near the valence band maximum for $\Delta_1=20$~meV plotted over a range of $-0.08<k_{x,y}a_0<0.08$. Contours span a range of energy of 10 meV.   
\textbf{e}, Resistivity  as a function of electron density $n_e$ and perpendicular displacement field $D$ measured at base temperature. 
Two disjoint regions showing signatures of superconductivity are observed, indicated by the open circles. 
\textbf{f-g}, Temperature and current dependence of the differential resistivity $dV/dI$ measured at the points in the $n-D$ plane indicated in panel a. 
\textbf{h} Temperature dependent resistivity across SC1 measured at $D=$0.46V/nm.
\textbf{i} $R_{xx}(T)$ with $T_{1/2}$ and $T_\mathrm{BKT}$ corresponding to the data plotted in panel f.
}\label{fig:fig1}
\end{figure*}

\textbf{
We report the observation of superconductivity in rhombohedral trilayer graphene electrostatically doped with holes.  
Superconductivity occurs in two distinct regions within the space of gate-tuned charge carrier density and applied electric displacement field, which we denote SC1 and SC2.  
The high sample quality allows for detailed mapping of the normal state Fermi surfaces by quantum oscillations, which reveal that in both cases superconductivity arises from a normal state described by an annular Fermi sea that is proximal to an isospin symmetry breaking transition where the Fermi surface degeneracy changes\cite{zhou_half_2021}. 
The upper out-of-plane critical field $B_{C\perp}\approx 10 \mathrm{mT}$ for SC1 and $1\mathrm{mT}$ for SC2, implying\cite{tinkham_michael_introduction_1975} coherence lengths $\xi$ of 200nm and 600nm, respectively.  The simultaneous observation of transverse magnetic electron focusing\cite{taychatanapat_electrically_2013,lee_ballistic_2016} implies a mean free path $\ell\gtrsim3.5\mathrm{\mu m}$.  Superconductivity is thus deep in the clean limit, with the disorder parameter\cite{abrikosov_contribution_1960} $d=\xi/\ell<0.1$. 
SC1 emerge from a paramagnetic normal state, and is suppressed with in-plane magnetic fields in agreement with the Pauli paramagnetic limit\cite{clogston_upper_1962,chandrasekhar_note_1962}. 
In contrast, SC2 emerges from a spin-polarized, valley-unpolarized half-metal\cite{zhou_half_2021}.  Measurements of the in-plane critical field show that this superconductor exceeds the Pauli limit by at least one order of magnitude.
We discuss our results in light of several mechanisms including conventional phonon-mediated pairing\cite{bardeen_theory_1957,gorkov_superconducting_2016}, pairing due to  fluctuations of the proximal isospin order\cite{scalapino_common_2012}, and intrinsic instabilities of the annular Fermi liquid\cite{kohn_new_1965,chubukov_superconductivity_2017}. 
Our observation of superconductivity in a clean and structurally simple two dimensional metal hosting a variety of gate tuned magnetic states may enable a new class of field-effect controlled mesoscopic electronic devices combining correlated electron phenomena.  
}

Owing to the instability of Fermi liquids to arbitrarily weak attractive interactions\cite{cooper_bound_1956}, most elemental metals become superconducting at sufficiently low temperatures. However, some metals become magnetic instead.  In these systems, time reversal symmetry is spontaneously broken, suppressing conventional superconducting pairing that relies on the degeneracy of Kramers pairs. 
The competition between magnetism and superconductivity can be understood from the point of view of the density of states: high density of states simultaneously favors superconductivity\cite{bardeen_theory_1957} and magnetism\cite{stoner_collective_1938}, with the ground state determined by the relative strength of the effective attractive interaction---typically mediated by phonons---and inter-electron Coulomb repulsion.
In other situations, for example in heavy-fermion compounds\cite{mathur_magnetically_1998}, 
magnetism and superconductivity may be cooperative.  In this scenario, magnetic fluctuations may themselves mediate attractive interactions between electrons\cite{scalapino_common_2012}, typically resulting in superconductivity with pairing symmetries other than s-wave. 

Here we report the discovery of superconductivity in rhombohedral trilayer graphene (RTG) on the cusp of an isospin symmetry breaking transition.
The crystal structure of RTG is shown in Fig. \ref{fig:fig1}a.  
As in other honeycomb carbon systems, near zero doping the Fermi surfaces are localized to the two inequivalent valleys at the corners of the hexagonal Brillouin zone.  Of relevance for isospin symmetry breaking, the valley provides an internal degree of freedom in addition to the electron spin. In the absence of an applied perpendicular displacement field $D$, the nonineracting electronic structure\cite{koshino_trigonal_2009,zhang_band_2010} of RTG is described by three Dirac crossings in each valley (see Fig. \ref{fig:fig1}b). 
At finite carrier density $n_e\approx 10^{12} cm^{-2}$, these Dirac pockets merge at a saddle-point van Hove singularity where the density of states diverges (\ref{fig:fig1}c-d).  
At finite $D$ the Dirac cones become gapped, and the van Hove singularities are enhanced in magnitude.  
Experimentally, RTG hosts a cascade of transitions at finite doping\cite{lee_gate_2019,zhou_half_2021} where one or more of the spin- and valley symmetries spontaneously breaks. These instabilities appear to be generic to rhombohedral graphite\cite{shi_electronic_2020}, and are predicted to apply to Bernal bilayer graphene as well\cite{castro_low-density_2008}.

\subsection{Superconducting phenomenology in RTG}
Our main result is summarized in Fig. \ref{fig:fig1}e, which shows a false-color plot of the longitudinal resistivity $R_{xx}$ as a function of $D$ and $n_e$. 
We observe two distinct superconducting states at these densities, which render in bright cyan on the color scale of Fig. \ref{fig:fig1}e and which we denote SC1 and SC2. 
 Both states show nonlinear transport signatures typical of superconductivity at sufficiently low temperatures (see Figs. \ref{fig:fig1}f-g and additional data in Fig. \ref{fig:S:dome_main_sc}). 
Fitting the nonlinear voltage to a Berezinskii–Kosterlitz–Thouless model\cite{berezinskii_destruction_1970,kosterlitz_ordering_1973} gives $T_\mathrm{BKT}=106\mathrm{mK}$ for SC1, while for SC2 $T_\mathrm{BKT}$ appears to be just below the base temperature of our measurement system (see Fig. \ref{fig:fig1}j and \ref{fig:S:bkt_2sc}). 

\begin{figure}[t]
\centering
\includegraphics[width=8.9cm]{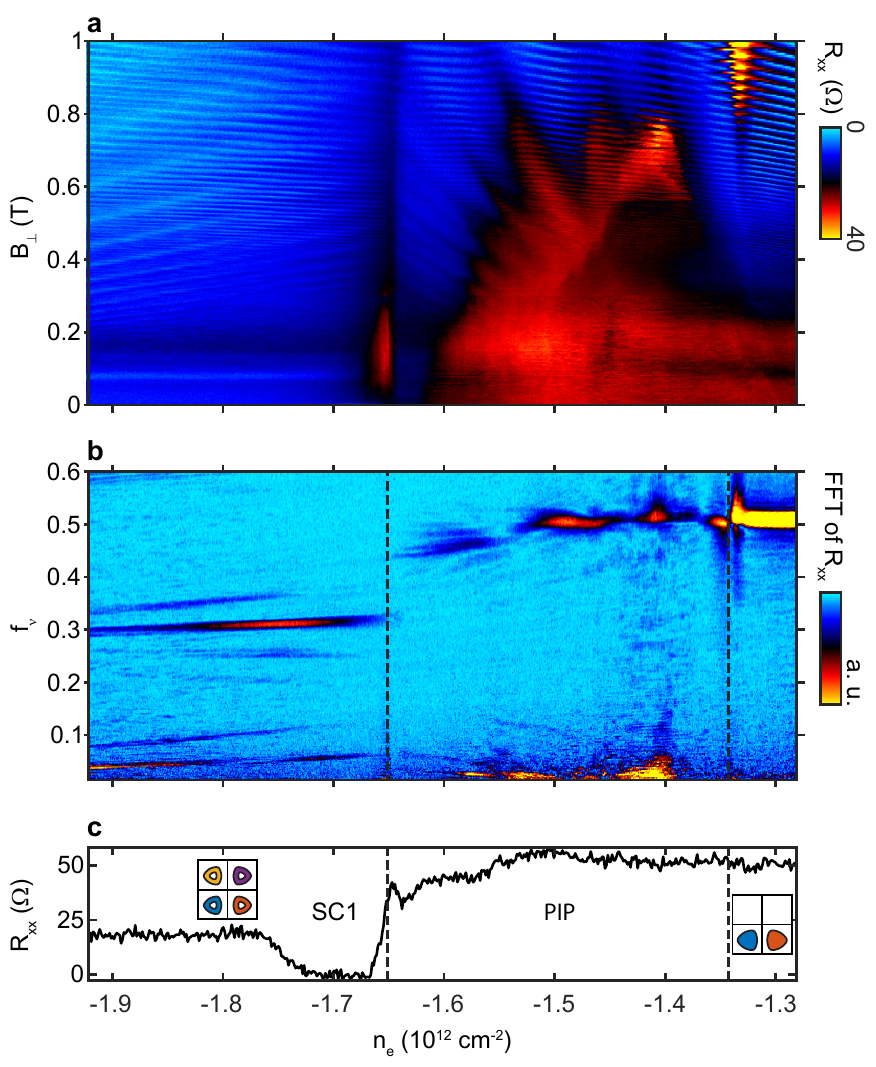}
\caption{\textbf{Fermiology of SC1.}
\textbf{a}, $R_{xx}$ vs $n_{\rm e}$ and $B_\perp$ measured at $D=$0.4V/nm.
\textbf{b}, Fourier transform of $R_{xx}(1/B_\perp)$, plotted as a function of frequency $f_\nu$ defined int he main text.  Dashed lines demarcate regimes of different Fermi surface topology. 
\textbf{c}, $R_{xx}$ vs $n_{\rm e}$ at $B=$0. Insets: Schematic of Fermi contours of the different phases identified from the quantum oscillations.
}\label{fig:fig2}
\end{figure}

Notably, both superconducting states occur near transitions in the normal state resistivity associated with a change in the degeneracy of the Fermi surface---in other words, superconductivity occurs at a symmetry breaking  transition. 
To better understand this connection, we measure quantum oscillations at low magnetic fields $B_\perp<1T$ (Fig. \ref{fig:fig2}a) in the density range spanning SC1 at fixed $D=0.4$V/nm.  Several oscillation periods are visible across this range, indicating complex Fermi surfaces. 
To understand these data more quantitatively, we plot the Fourier transform of $R_{xx}(1/B_\perp)$ as a function of $f_\nu$, the oscillation frequency normalized to the total carrier density (Fig. \ref{fig:fig2}b). 
$f_\nu$ corresponds to the fraction of the total Fermi sea area enclosed by the Fermi surface generating the peak. 
Three regions of qualitatively different quantum oscillation spectra are visible. 
At extreme right, a single peak at $f_\nu=.5$ indicates two equal area Fermi surfaces each enclosing half the total Fermi sea.  We associate this regime with a spin polarized, valley unpolarized ``half-metal'  state\cite{zhou_half_2021} with a simply connected Fermi sea in each valley.  
At the extreme left of the plot, several oscillation peaks with density dependent frequencies are visible.  These correspond to the inner- and outer boundaries of an annular Fermi sea with the full four-fold spin- and valley-degeneracy (and harmonics).  Intermediate between these two phases, the oscillation spectrum is more complex, including both strong peaks at $f_\nu\lesssim.5$ as well as at $f_\nu<.1$.  We identify this regime with one or more partially isospin polarized (PIP) phases, where the system has broken one of the spin- or valley symmetries but is not completely polarized into two isospin components.    
Comparing the quantum oscillation spectrum to base temperature transport measurements at B=0 (Fig. \ref{fig:fig2}c) shows that SC1 occurs within the symmetric, annular phase and adjacent to the boundary with the PIP phase.  

\begin{figure}[b]
\centering
\includegraphics[width=\columnwidth]{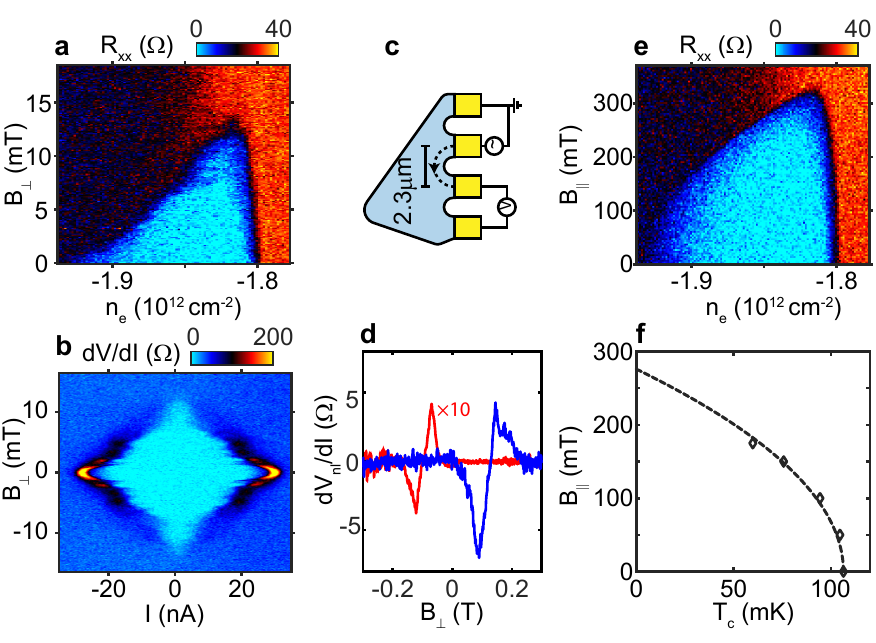}
\caption{\textbf{Magnetic field dependence of SC1.}
\textbf{a}, $B_\perp$-dependent $R_{xx}$  measured at $D=$0.4V/nm.
\textbf{b}, $B_\perp$-dependence of the nonlinear resistivity at $n_e=-1.83\times10^{12}\rm{cm}^{-2}$ and $D=$0.46V/nm.
\textbf{c}, Device and circuit schematic for measuring transverse magnetic electron focusing.
\textbf{d}, Magnetic-focusing induced non-local voltage measured at $n_e=-1.83\times10^{12}\rm{cm}^{-2}$, $D=$0.46V/nm (blue) and $n_e=1.83\times10^{12}\rm{cm}^{-2}$, $D=$0 (red). For the red curve, the nonlocal resistivity has been multiplied by 0.1.
\textbf{e}, $B_{||}$-dependent $R_{xx}$ measured at $D=$0.4V/nm.
\textbf{f}, $B_{\parallel C}$ dependence of $T_\mathrm{BKT}$ and $T_{1/2}$ measured at $n_e=1.83\times10^{12}\rm{cm}^{-2}$, $D=$0.46V/nm (see also Fig. \ref{fig:S:BKTmain}). Lines show fits to the phenomenological relation $T_C/T_C^0=1-\left(B_{C \parallel}/B_{C \parallel}^0\right)^2$. 
}\label{fig:fig3}
\end{figure}

The appearance of superconductivity so close to a symmetry breaking phase transition opens the possibility of an unconventional superconducting state. A characteristic of many unconventional superconductors is their fragility with respect to disorder, due to the inapplicability of Anderson's theorem\cite{anderson_theory_1959}.  Disorder in superconductors is quantified by the ratio of the coherence length ($\xi$) to the mean free path ($\ell$), $d=\xi_0/\ell$, with the superconductivity destroyed when $d\approx1$ for unconventional superconductors\cite{abrikosov_contribution_1960}. 
To assess $d$ in RTG, we study the magnetoresistance of both the superconducting and normal states.  
Fig. \ref{fig:fig3}a-b show the dependence of SC1 on the out-of-plane magnetic field $B_\perp$.  The critical $B_{C\perp}$ is in the 10mT range.  Within Ginzburg-Landau theory, $B_{C\perp}$ is related\cite{tinkham_michael_introduction_1975} to the coherence length by $2 \pi\xi^2=\phi_0/B^C_\perp$, where $\phi_0$ is the superconducting flux quantum.  As a result, $\xi\approx 150-250 nm$ for SC1. 
$\ell$ may be estimated from the Drude conductivity $R\approx \frac{h}{e^2} \frac{1}{4 k_f \ell}$ where $h$ is Planck's constant, $e$ is the elementary charge, $k_f$ is the Fermi wave vector. Taking $k_f=\sqrt{\pi n_e}\approx .25 nm^{-1}$ and a normal state resistance of $R\approx 20 \Omega$ produces an estimate of $\ell\approx 1\mu m$, considerably larger than $\xi$ and implying $d\lesssim .2$. 
However, this estimate for $\ell$ is comparable to the lateral dimensions of our device (Fig. \ref{fig:fig3}c), calling into question the validity of the Drude approach\cite{wang_one-dimensional_2013}. In fact, qualitative features suggest $\ell$ may be considerably longer.  Fig. \ref{fig:fig3}c-d show a circuit schematic for measuring the nonlocal magnetoresistance, which has been used to detect transverse electron focusing in other graphene heterostructures\cite{taychatanapat_electrically_2013,lee_ballistic_2016}. Measured data in the regime of SC1 (Fig. \ref{fig:fig3}d) show a pronounced feature near $B_\perp\approx.1T$, consistent with transverse electron focusing between the contacts, which are separated by a pitch of $L\approx 2.3\mu m$. This feature--which is observed across all densities in our device (Fig. \ref{fig:S:focusing})---suggests $\ell\gtrsim \pi L\approx 3.5\mu m$.  Taking this estimate for $\ell$ gives a disorder parameter $d<.035$.  These estimates place the superconductivity firmly in the clean limit, where unconventional superconductivity may be expected to survive. 

To further explore the properties of SC1, we show the response to an in-plane magnetic field in Fig. \ref{fig:fig3}e. The in-plane critical field $B_{\parallel C}$ is several hundred millitesla, more than one order of magnitude larger than $B_{\perp C}$ consistent with the 2D nature of the superconductivity. To explore the mechanism for the magnetic field induced breakdown of superconductivity, Fig. \ref{fig:fig3}f shows the dependence of  both $T_\mathrm{BKT}$ and $T_{1/2}$ on $B_\parallel$.  
The data are well fit by the relation $T_C/T_C^0=1-\left(B_{C \parallel}/B_{C \parallel}^0\right)^2$ for a superconductor limited by Pauli paramagnetism\cite{clogston_upper_1962,chandrasekhar_note_1962}, where  $B_{C\parallel}^0$ and $T_{C}^0$ describe the T=0 critical field and $B_\parallel=0$ critical temperature, respectively. 
For both fits, we find $\mu_B B_{C\parallel}^0/(k_B T_C^0)=1.7$, close to the values 1.23 predicted by weak coupling BCS theory without accounting for the Coulomb repulsion or finite temperature effects. We thus conclude that the $B_\parallel$ dependence is likely compatible with a conventional spin-singlet order parameter.  

\begin{figure}[t]
\centering
\includegraphics[width=\columnwidth]{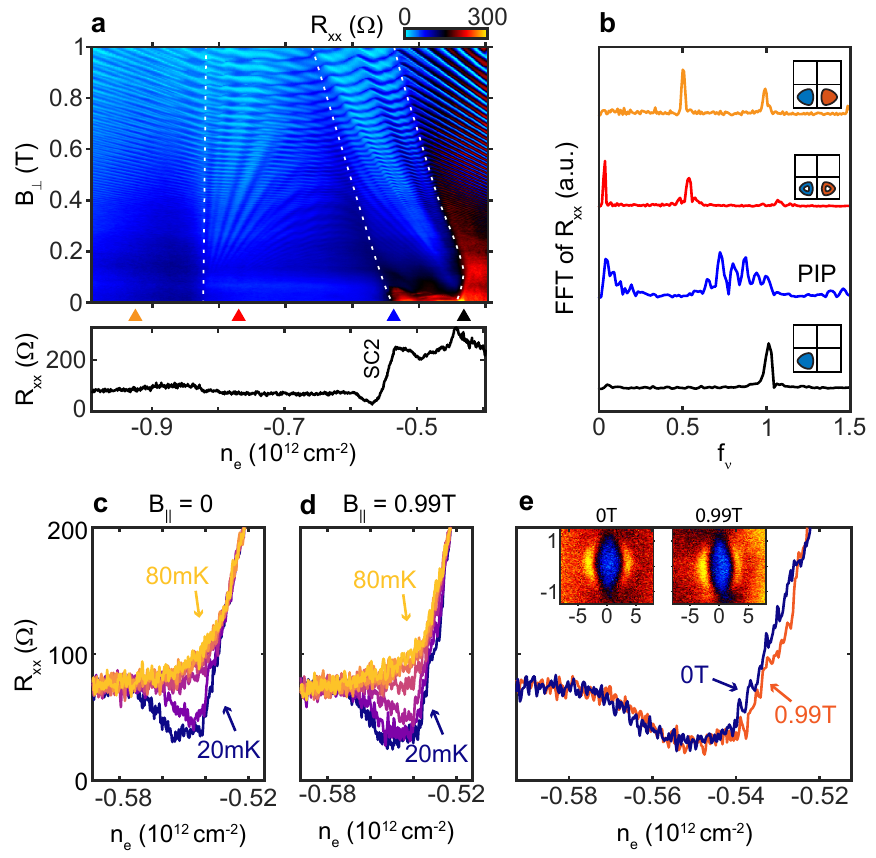}
\caption{\textbf{Fermiology and $B_\parallel$ dependence of SC2.}
\textbf{a}, Top: $R_{xx}$ as a function of $n_{\rm e}$ and $B_\perp$ for $D=$0.33V/nm. 
Bottom: $R_{xx}$ at $B=0$ for $D=$0.33V/nm.
\textbf{b} Fourier transforms of $R_{xx}(1/B_\perp)$ for the values of $n_e$ indicated by arrows in panel a. Insets: Schematic Fermi contours.
\textbf{c}, Temperature dependence of $R_{xx}$ vs $n_{\rm e}$ measured at $D=$0.33V/nm.
\textbf{d}, Same as c, measured with an 0.99T in-plane magnetic field applied.
\textbf{e}, In-plane magnetic field dependence of $R_{xx}$ vs $n_{\rm e}$ measured at $D=$0.33V/nm. Insets: $dV/dI$ measured at $n_e=-1.83\times10^{12}\rm{cm}^{-2}$ and $D=$0.46V/nm. x-axis is $I$(nA), y-axis is $B_\perp$(mT).
}\label{fig:fig4}
\end{figure}

In contrast, the phenomenology of SC2 is not compatible with conventional spin-singlet pairing. 
As is evident from the  quantum oscillations shown in Figs. \ref{fig:fig4}a-b, SC2 emerges from a two-fold degenerate annular Fermi sea associated with a spin-polarized half-metal\cite{zhou_half_2021}.   
While the low $T_C$ of SC2 complicates quantitative analysis of the kind presented for SC1, signatures of SC2 persist to very large values of $B_\parallel$, with $B_{C\perp}$ and the critical current nearly unchanged for $B_\parallel$ as high as 1T (Figs. \ref{fig:fig4}c-e and Fig. \ref{fig:s:triplet}).
Taking a conservative estimate of $50\mathrm{mK}$ for $T_C^0$ and 
$1T$ for $B_{C\parallel}^0$, SC2 violates the Pauli limit by more than one order of magnitude, consistent with a spin-polarized superconductor.

For attractive interactions of finite range, such as arise from electron phonon interactions, pairing potentials are attractive in all angular momentum channels. 
The potential is strongest in the s-wave channel, favoring conventional spin-singlet pairing in normal metals.  
In the spin-polarized half-metal regime where SC2 occurs, electrons with reversed spin are separated energetically from the ground state by the exchange energy, which at several meV\cite{zhou_half_2021} is at least two orders of magnitude larger than observed superconducting gaps. Spin-singlet pairing is thus energetically precluded. 
The unique properties of graphene nevertheless allow for superconductivity from conventional pairing mechanisms.  
Most importantly, the negligible spin-orbit coupling endows the spin-polarized half-metal with spinless time-reversal symmetry, which guarantees degeneracy between electrons in opposite valleys but with the same spin even in the absence of inversion symmetry.  
The smaller $T_C$ of SC2 relative to SC1 is consistent with pairing in a higher angular momentum channel by the same interaction.  One natural order parameter, proposed for moire systems with similar symmetries, is the spin-triplet, valley singlet  $\langle\hat c^\dagger_{k\uparrow}\hat c^\dagger_{-k,\uparrow}\rangle$\cite{lee_theory_2019,cornfeld_spin-polarized_2021}. 
This form of superconductivity shares many similarities with conventional superconductors, most notably protection from intra-valley scattering by smooth disorder potentials.  

\section{Discussion}

The common features shared by SC1 and SC2 suggest several possible mechanisms, both conventional and all-electronic. 

Most obviously, the appearance of superconductivity near symmetry breaking phase transitions suggests that fluctuations of the proximal ordered state may play a role in pairing\cite{scalapino_common_2012}.  
The plausibility of this picture hinges on the nature of the transition.  Experimentally, the sudden jump in quantum oscillation spectra observed near the superconductors is suggestive of a first order transition. In this case, fluctuations might be suppressed. 
However, the resistivity of the normal state changes only gradually across the transition, contrasting with other isospin transitions studied in the same sample that are strongly first order\cite{zhou_half_2021}. Measurements of the thermodynamic compressibility\cite{zhou_half_2021} similarly do not show strong negative compressibility where superconductivity is observed, allowing for the possibility of a continuous transition.  

The nature of the proximal ordered state also plays a key role in fluctuation mediated superconductors, with different orders producing attraction in different pairing channels\cite{scalapino_common_2012}. 
In RTG, in-plane field measurements show that the PIP phase proximal to SC1 is likely spin-unpolarized (Fig. \ref{fig:S:proximalphase}). 
To match the experiment, then, a  theory of fluctuation mediated superconductivity for SC1 should produce an apparently Pauli-limited superconductor from fluctuations of a spin-unpolarized isospin ordered state--a strong constraint. 

Alternatively, superconductivity and symmetry breaking may arise in close proximity from unrelated mechanisms.  Within BCS theory, the superconducting transition temperature in the antiadiabatic limit\cite{gorkov_phonon_2016} applicable to low-density electron systems is approximated by 
\begin{align}
    T_C&=T_F e^{-1/\lambda}\label{TC}
\end{align}
where the Fermi temperature $T_F\approx 50K$ in the regime of interest and $\lambda=g\rho$ is the dimensionless coupling constant characterizing attractive interactions, which depends on the coupling constant $g$ and the density of states.  
For a density independent$g$---as expected for  phonon mediated attraction, for example----superconductivity is observed at temperature $T$ when $\rho$ exceeds $\rho_{SC}=\frac{1}{g\log(T_F/T)}$. This approach has been used to predict superconductivity in rhombohedral graphite\cite{kopnin_surface_2014}.
However, high density of states also favors symmetry breaking, with the boundary between ordered and disordered states defined by the Stoner criterion, $\rho_{FM}>1/U$ where $U$ parameterizes the Coulomb repulsion.  

As $\rho$ increases---as occurs in our experiment as $|n_e|$ is reduced---one of two scenarios obtains. 
For $\rho_{FM}<\rho_{SC}$, the Stoner criterion is satisfied first, and the Fermi liquid becomes magnetic. As a result, $\rho$ decreases, the Kramers degeneracy is lifted, and superconductivity is not observed.  
Conversely, if $\rho_{FM}>\rho_{SC}$, superconductivity is observed.  However, as the density of states is \textit{further} increased above $\rho_{FM}$, the system nevertheless becomes magnetic.  In this case the domain of superconductivity is  bounded from below by $\rho_{SC}$ and from above by $\rho_{FM}$. Superconductivity occurs at the cusp of a magnetic transition, precisely as observed, despite the lack of a causal link between the two. 

Bolstering the case for this scenario is the fact that both superconductors arise at the threshold of a magnetic transition but are predominantly within the \textit{disordered phase}; quantities such as $B_{C\parallel}$ and $B_{C\perp}$ rise gradually as the isospin symmetry breaking transition is approached before rapidly collapsing at the transition itself. However, a key question remains as to whether this picture is consistent with the seemingly narrow range of $n_e$ over which superconductivity is observed. 
For example, SC1 occurs over a density range $\Delta n/n\approx 5\%$. Comparing the maximum $T_C\approx 100\mathrm{mK}$ to our estimated base temperature of 30-40mK, we estimate $\Delta T_C/T_C\approx .6-.7$ over this same range. For this to be accounted for entirely by a change in $\lambda$, $\Delta\lambda/\lambda\approx .1$, about twice as large as expected from single-particle calculations of the density of states.  More detailed calculations (for instance, accounting for both the Coulomb repulsion and finite temperature effects\cite{mcmillan_transition_1968}) may assess whether this quantitative discrepancy is significant. 

In both phonon- and fluctuation-mediated superconductors, high temperature transport typically shows signs of electron scattering by the same neutral modes that mediate the superconductivity\cite{allen_electron-phonon_2000}. 
We find no sign of enhanced high temperature scattering, at least up to 20K (Fig. \ref{fig:S:RvsT}). A mechanism for superconductivity---albeit not usually in the  s-wave channel---that does not invoke soft modes was given by Kohn and Luttinger based on the intrinsic instability of the Fermi liquid\cite{kohn_new_1965}.  While thought to occur only at experimentally inaccessible temperatures and disorder strengths in most materials, it has been proposed\cite{chubukov_superconductivity_2017} that in semiconductor quantum wells with two occuppied subbands, this effect may be enhanced. Given the similarity between a two subband system and the annular Fermi seas we describe above, combined with the exceptionally low disorder in RTG, exploration of such mechanisms may be warranted. 

In closing, we comment on the possible relationship between the superconductivity reported here and that observed in moir\'{e} systems. 
In RTG aligned to hexagonal boron nitride, the moire potential only weakly perturbs the underlying isospin symmetry breaking\cite{zhou_half_2021}. The $n_e$ and $B_\parallel$ dependence of the signatures of superconductivity observed in that system\cite{chen_signatures_2019} would appear to be most consistent with SC2.  
Twisted bilayer\cite{cao_unconventional_2018} and twisted trilayer\cite{park_tunable_2021,hao_electric_2021} have different microscopic symmetries; however, they share several features with RTG including enhanced density of states  and isospin symmetry breaking. We conjecture that the superconductivity observed in all graphene systems has the same basic origin.

\section{Methods}
The trilayer graphene and hBN flakes were prepared by mechanical exfoliation of bulk crystals. The rhombohedral domains of trilayer graphene flakes were detected using a Horiba T64000 Raman spectrometer with a 488nm mixed gas Ar/Kr ion laser beam. The rhombohedral domains were subsequently isolated using anodic oxidation cutting with an atomic force microscope\cite{masubuchi_fabrication_2009}. The Van der Waals heterostructures were fabricated following a dry transfer procedure\cite{wang_one-dimensional_2013}, with care taken to minimize the mechanical stretching of rhombohedral trilayer graphene. Fabrication details are described in \onlinecite{zhou_half_2021}, which studied the same device. 

Transport measurement was performed using a lock-in amplifier. Data in Fig. \ref{fig:S:dome_main_sc} were measured at a frequency of 19.177Hz. The rest data were measured at 42.5Hz. The frequency was chosen to minimize electronic noise.

All measurements were performed in a dilution refrigerator equipped with a vector superconducting magnet. Unless specified, measurements were performed at base temperature, corresponding to $T\lesssim$20mK as measured by a calibrated Ruthenium oxide thermometer mounted close to the sample. Cryogenic low-pass filters are applied to reduce the electron temperature.

\section*{acknowledgments}
The authors acknowledge extensive discussions with E. Berg and M. Zaletel, and thank them for their comments on the completed manuscript. The authors also acknowledge discussions with S. Kivelson, S. Das Sarma, A. Bernevig, and A.H Macdonald. We acknowledge experimental assistance from L. Cohen, who installed the dilution refrigerator. 
This project was primarily funded by the Department of Energy under DE-SC0020043. AFY acknowledges the support of the Gordon and Betty Moore Foundation under award GBMF9471. 
K.W. and T.T. acknowledge support from the Elemental Strategy Initiative conducted by the MEXT, Japan, Grant Number JPMXP0112101001 and JSPS KAKENHI, Grant Number JP20H00354.  

\section*{Author contributions}
HZ fabricated the device with assistance from TX.    HZ performed the measurements, advised by AFY.    KW and TT grew the hexagonal boron nitride crystals.    HZ, and AFY wrote the manuscript with input from all authors.

\normalem
\let\oldaddcontentsline\addcontentsline
\renewcommand{\addcontentsline}[3]{}
\bibliography{references}
\let\addcontentsline\oldaddcontentsline

\renewcommand\thefigure{S\arabic{figure}}
\setcounter{figure}{0}

\clearpage
\onecolumngrid
\begin{center}
\textbf{\large Supplementary information for ``Superconductivity in rhombohedral trilayer graphene'' }\\[5pt]
\begin{quote}
 {\small 
}
\end{quote}
\end{center}
\setcounter{equation}{0}
\setcounter{table}{0}
\setcounter{page}{1}
\setcounter{section}{0}
\makeatletter
\renewcommand{\theequation}{S\arabic{equation}}
\renewcommand{\thefigure}{S\arabic{figure}}
\renewcommand{\thepage}{S\arabic{page}}

\begin{figure*}[h]
\centering
\includegraphics[width=\textwidth]{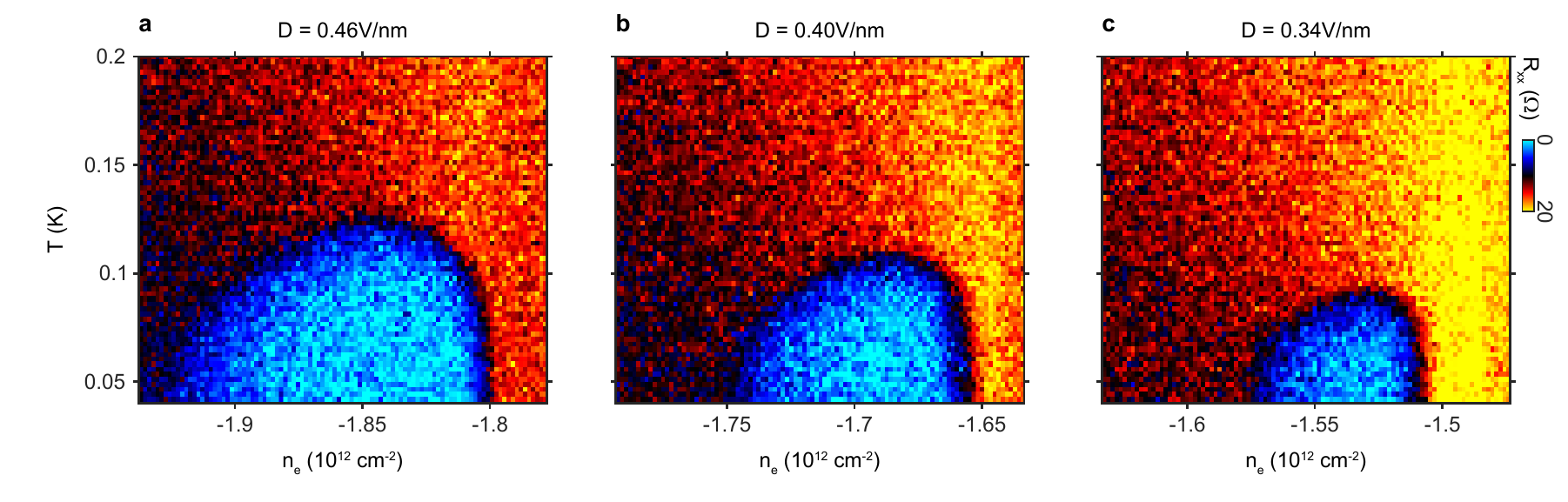}
\caption{\textbf{Displacement field dependence of SC1.}
\textbf{a}, $R_{xx}$ as a function of $n_{\rm e}$ and $T$ at $D=$0.46V/nm;
\textbf{b}, at $D=$0.40V/nm;
\textbf{c}, at $D=$0.34V/nm.
}\label{fig:S:dome_main_sc}
\end{figure*}

\begin{figure*}[h]
\centering
\includegraphics[width=12cm]{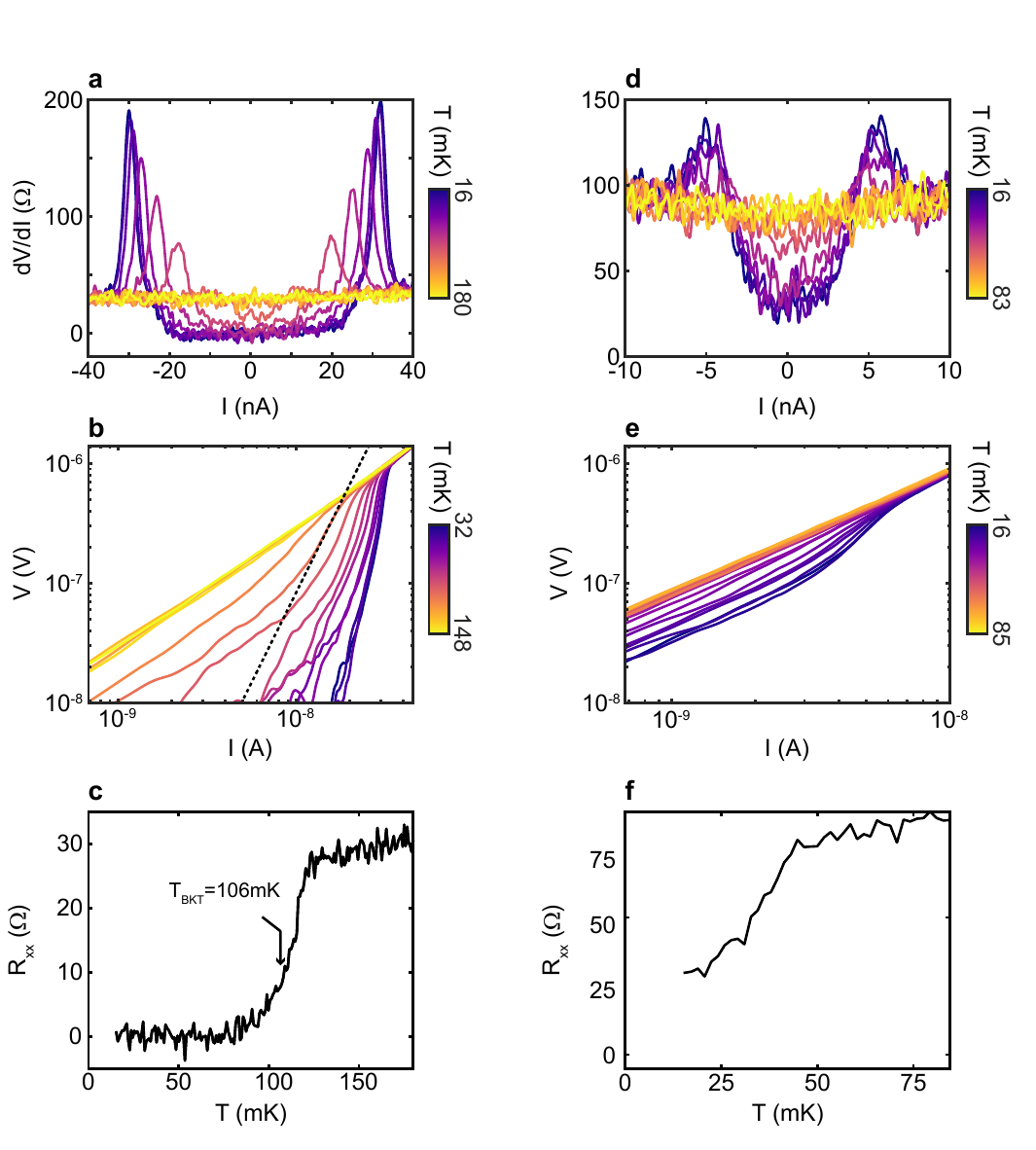}
\caption{
\textbf{Temperature dependent data for SC1 and SC2.}
\textbf{a}, Temperature dependent $dV/dI$ measurements of SC1. Measurements were performed at $n_e=-1.8\times10^{12}\rm{cm}^{-2}$, $D=$0.46V/nm.
\textbf{b}, $V(I)$ for SC1.  The dashed line shows $V\propto I^3$; we take $T_{BKT}$ as the highest temperature where the $V(I)$ curve shows $I^3$ scaling. 
\textbf{c}, 
$R_{xx}(T)$ for SC1 with $T_{BKT}$ indicated.
\textbf{d}, Same as panel a, but for SC2. Measurements were performed at $n_e=-0.55\times10^{12}\rm{cm}^{-2}$, $D=$0.33V/nm.  
\textbf{e}, Same as panel b, but for SC2.
\textbf{f}, Same as panel c, but for SC2.
}\label{fig:S:bkt_2sc}
\end{figure*}

\begin{figure*}[b]
\centering
\includegraphics[width=\textwidth]{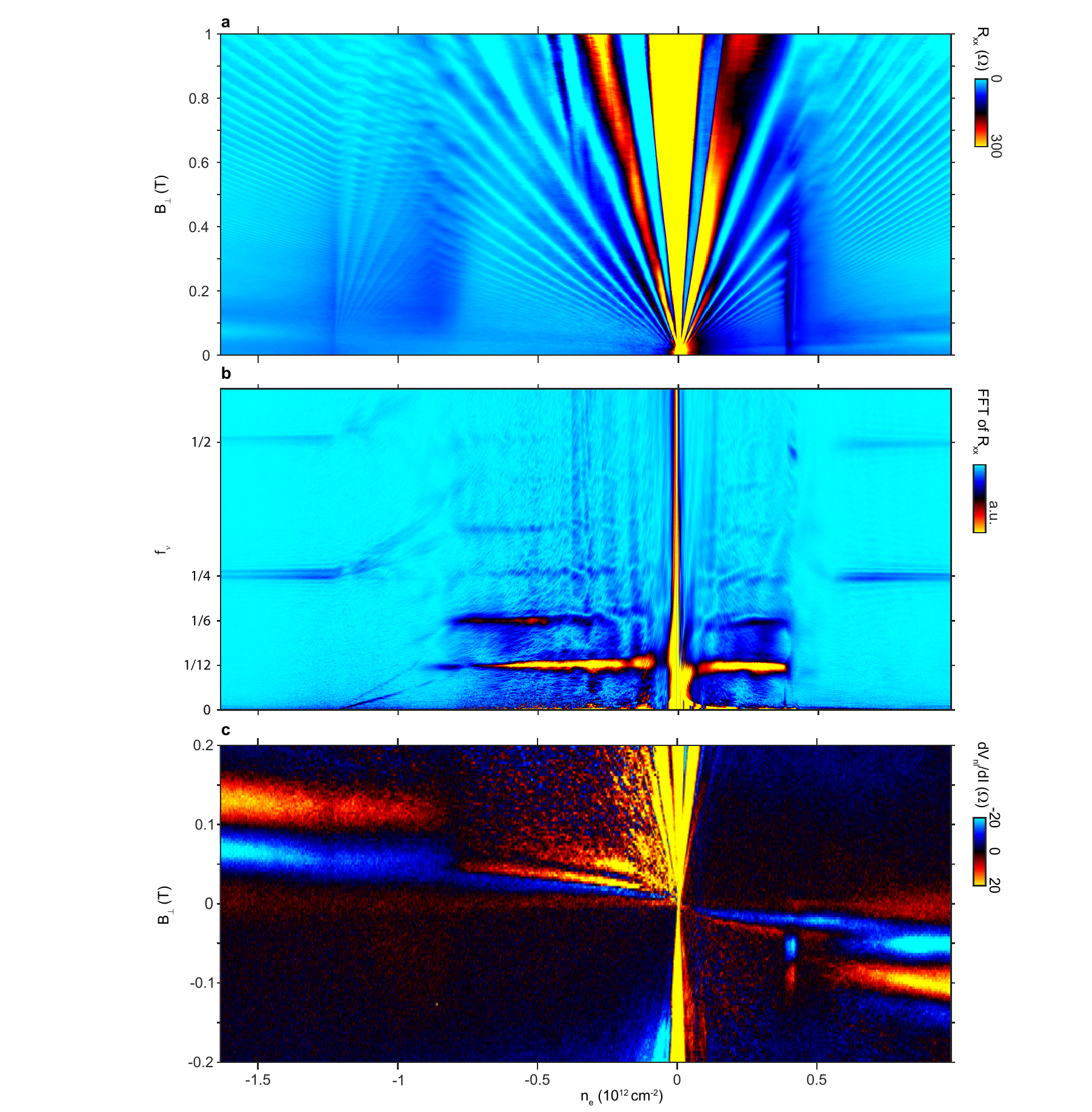}
\caption{\textbf{Comparison of quantum oscillations and transverse magnetic electron focusing at $D=$0.} 
\textbf{a}, $R_{xx}$ vs $n_{\rm e}$ and $B_{\perp}$ measured at $D=$0.
\textbf{b}, Fourier transform of $R_{xx}(1/B_{\perp})$ for data in panel a.
\textbf{c}, Non-local resistance measured in the configuration in Fig.\ref{fig:fig3}c as a function of $n_{\rm e}$ and $B_{\perp}$.
}\label{fig:S:focusing}
\end{figure*}

\begin{figure*}[b]
\centering
\includegraphics[width=\textwidth]{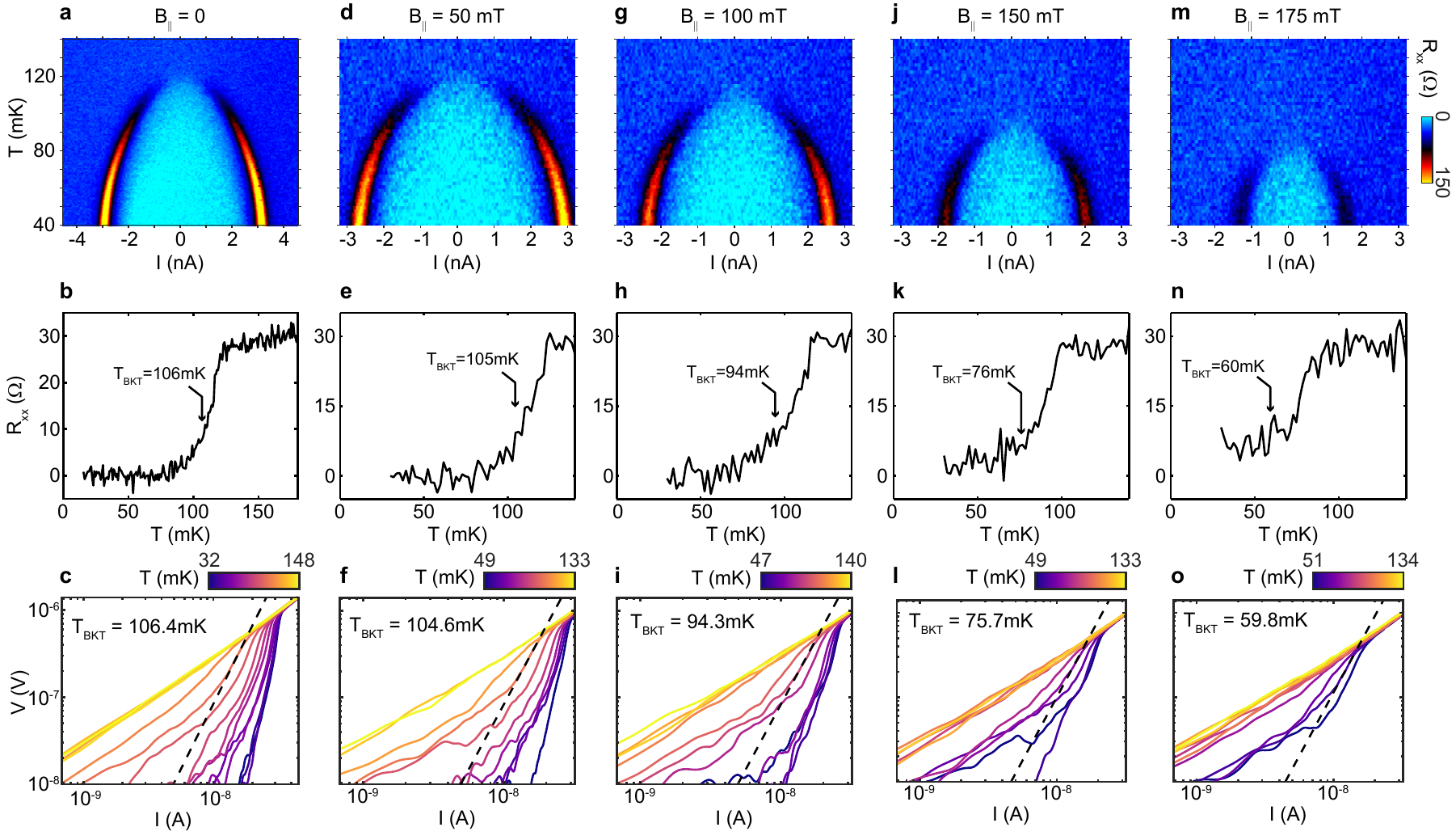}
\caption{\textbf{$B_\parallel$ dependence of SC1.}
\textbf{a}, Temperature dependent $dV/dI$ measurements of SC1. Measurements were performed at $n_e=-1.8\times10^{12}\rm{cm}^{-2}$, $D=$0.46V/nm.
\textbf{b}, $R_{xx}(T)$ for SC1 with $T_{BKT}$ indicated.
\textbf{c}, $V(I)$ for SC1.  The dashed line shows $V\propto I^3$ we take $T_{BKT}$ as the highest temperature where the $V(I)$ curve shows $I^3$ scaling.
\textbf{d-f}, Same as panel a-c measured at $B_{||}=50$mT.
\textbf{g-i}, Same as panel a-c measured at $B_{||}=150$mT.
\textbf{m-o}, Same as panel a-c measured at $B_{||}=175$mT.
}\label{fig:S:BKTmain}
\end{figure*}

\begin{figure*}[b]
\centering
\includegraphics[width=\textwidth]{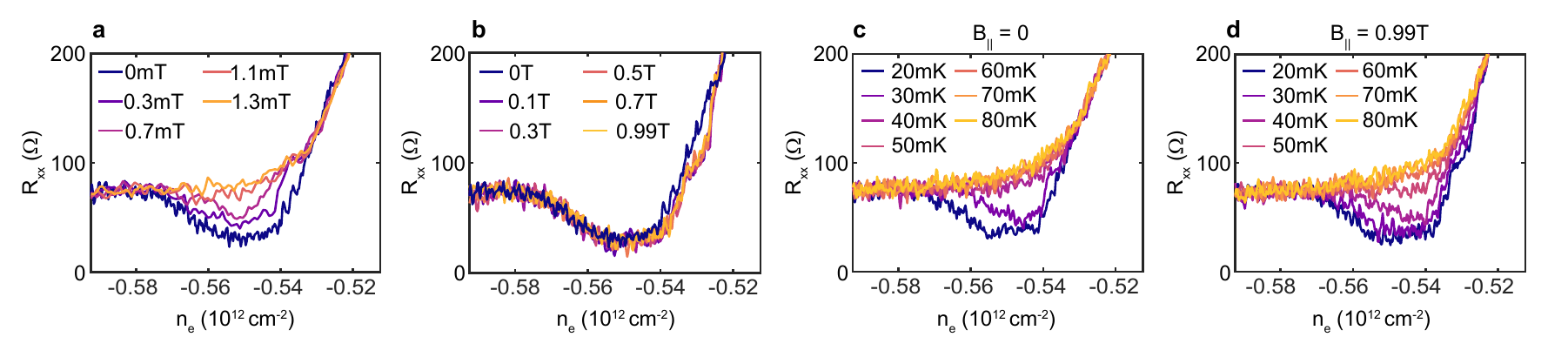}
\caption{\textbf{Magnetic field and temperature dependence of SC2}.
\textbf{a}, $\bm R_{xx}$ vs $n_{\rm e}$ measured at $D=$0.33V/nm and various $B_\perp$ with $B_\parallel=$0.
\textbf{b}, Same as a, measured at various $B_\parallel$ with $B_\perp$=0.
\textbf{c}, $\bm R_{xx}$ vs $n_{\rm e}$ measured at $D=$0.33V/nm and various temperature with $B_\perp=$0, $B_\parallel=$0
\textbf{d}, Same as c, with $B_\parallel=$0.99T.
}\label{fig:s:triplet}
\end{figure*}

\begin{figure*}[b]
\centering
\includegraphics[width=\textwidth]{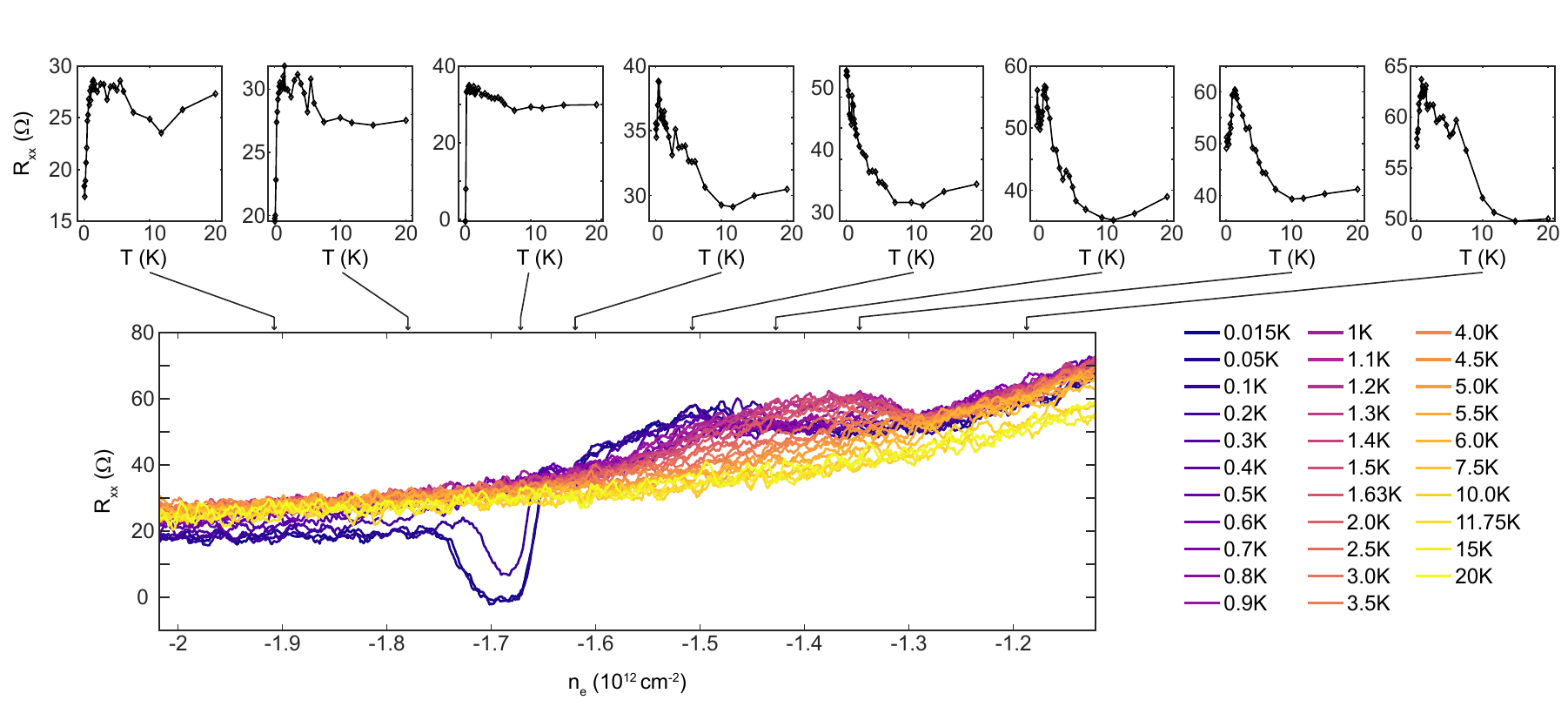}
\caption{\textbf{Temperature dependence of $\bm{R_{xx}}$ measured at $\bm{D=}$0.4V/nm and $\bm{n_{\rm e}<0}$
} Bottom panel shows $R_{xx}$ as a function of $n_{\rm e}$ at different temperature. Top panels show $R_{xx}$ vs $T$ at fixed $n_{\rm e}$ extracted from the bottom panel.
}\label{fig:S:RvsT}
\end{figure*}

\begin{figure*}[b]
\centering
\includegraphics[width=\textwidth]{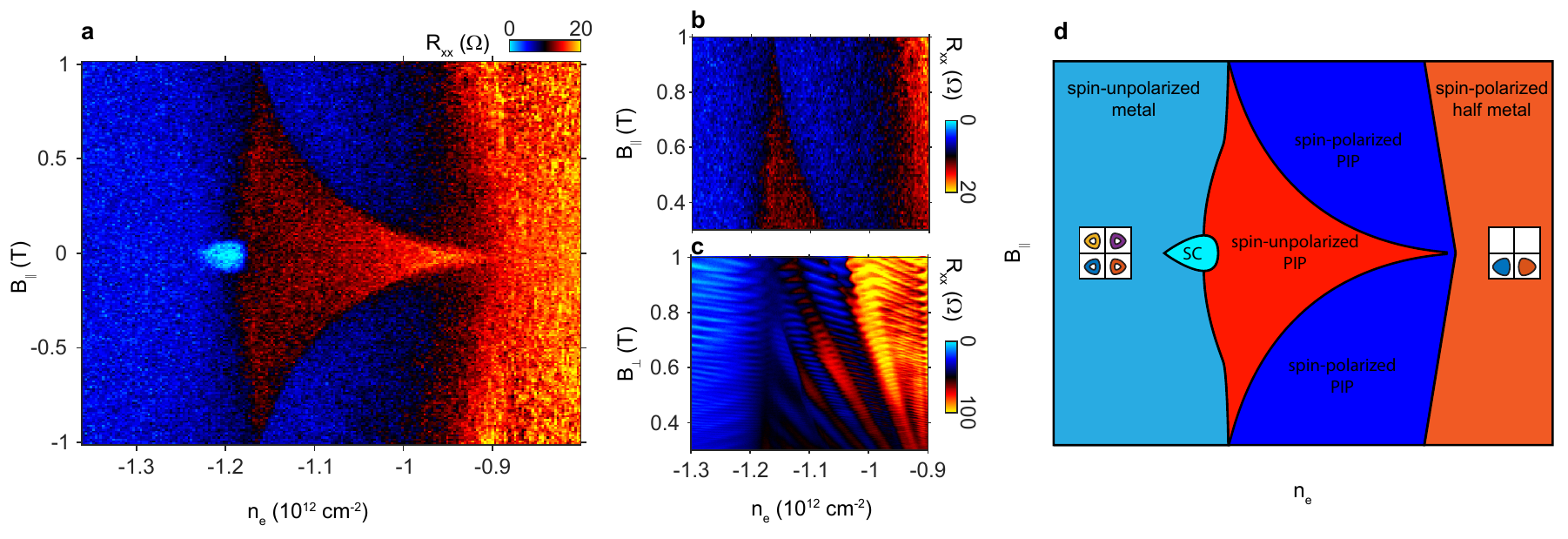}
\caption{\textbf{In-plane magnetic field dependence of the PIP phase near SC1.}
\textbf{a}, $B_{||}$ dependence of $R_{xx}$ near SC1 at $D=$0.228V/nm.
\textbf{b}, Zoom-in of panel a.
\textbf{c}, Same as panel b but measured with an out-of-plane field applied instead of in-plane field.
\textbf{d}, Schematic phase diagram extracted from panel a. Insets are schematic Fermi contours of the isospin polarized and unpolarized phases.
}\label{fig:S:proximalphase}
\end{figure*}
\end{document}